\documentclass[12pt,a4paper,reqno,draft]{amsart}
\usepackage{amsmath,amssymb,euscript,bbold}
\usepackage{latexsym}
\usepackage{array}
\renewcommand{\Sp}{\mathrm{Sp}}
\newcommand{\RR}{\mathbb{R}}
\newcommand{\sC}{\mathsf{C}}
\newcommand{\1}{\mathbb{1}}
\newcommand{\eL}{\EuScript{L}}
\newcommand{\eV}{\EuScript{V}}
\newcommand{\eW}{\EuScript{W}}
\newcommand{\eM}{\EuScript{M}}
\newcommand{\CS}{\mathsf{CS}}
\newcommand{\Spin}{\mathrm{Spin}}
\newcommand{\SO}{\mathrm{SO}}
\newcommand{\SU}{\mathrm{SU}}
\newcommand{\U}{\mathrm{U}}
\newcommand{\SL}{\mathrm{SL}}
\newcommand{\Cl}{\mathrm{C}\ell}
\DeclareMathOperator{\Tr}{\mathrm{Tr}}
\begin{document}

\title{Supersymmetry and gauge theory in Calabi--Yau 3-folds}
\author{JM Figueroa-O'Farrill}
\address{\begin{flushright}Department of Physics\\
Queen Mary and Westfield College\\
University of London\\
London E1 4NS, UK\end{flushright}}
\email{j.m.figueroa@qmw.ac.uk}
\thanks{Supported by the UK EPSRC under contract GR/K57824.}
\author{A Imaanpur}
\address{\begin{flushright}Department of Physics and Mathematical
Physics\\
University of Adelaide\\
Adelaide, SA 5005, Australia\end{flushright}}
\email{aimaanpu@physics.adelaide.edu.au}
\thanks{Supported by the Ministry of Culture and Higher Education,
Iran.}
\author{J McCarthy}
\email{jmccarth@physics.adelaide.edu.au}
\thanks{Supported by the Australian Research Council.}
\begin{abstract}
We consider the dimensional reduction of supersymmetric Yang--Mills on
a Calabi--Yau 3-fold.  We show by construction how the resulting
cohomological theory is related to the balanced field theory of the
K\"ahler Yang--Mills equations introduced by Donaldson and
Uhlenbeck--Yau.
\end{abstract}
\maketitle

\section{Introduction}

The study of Ricci-flat manifolds is interesting to both geometers and
string theorists for a variety of reasons. These manifolds provide
examples of ``exotic'' Einstein geometries: in fact, their holonomy
groups have to be $\SU(n)$, $\Sp(n)$, $G_2$ or $\Spin(7)$,
corresponding to Calabi--Yau $n$-folds, hyperk\"ahler manifolds of
real dimension $4n$, and exceptional 7- and 8-manifolds, respectively.
Because they are Ricci-flat and admit parallel spinors, they are
supersymmetric vacua for superstring-related theories.  Out of these
parallel spinors one can construct parallel forms \cite{Wang,H} which
turn out to be calibrations in the sense of \cite{HL}.  Indeed, these
manifolds have a rich geometry of (calibrated) minimal submanifolds.
These submanifolds are, in the simplest case, the supersymmetric
cycles \cite{BBS} around which branes may wrap to produce BPS states.
Yang--Mills theory on these manifolds is also interesting.  The
equations of motion admit instantonic solutions which minimise the
action and are defined by linear equations generalising
(anti)self-duality in four dimensions \cite{CDFN,Ward}.  This
observation forms the basis of the ``Oxford programme'' \cite{DT} to
generalise Donaldson--Floer--Witten theory to higher dimensional
Ricci-flat manifolds.

Perhaps one of the boldest proposals yet to have emerged out of the
``second superstring revolution'' is the Matrix Conjecture of Banks
{\em et al.\/} \cite{Matrix}.  This conjecture states that the
di\-men\-sion\-al reduction to one dimension of 10-di\-men\-sion\-al
supersymmetric Yang--Mills in the limit in which the rank of the gauge
group goes to infinity provides an Infinite Momentum Frame description
of M-theory, the 11-di\-men\-sion\-al theory believed to underlie
nonperturbative superstring theory.  In this context, it becomes an
important problem to understand the dimensional reductions on
10-dimensional supersymmetric Yang--Mills theory.  Most research has
focused on toroidal compactifications, since these preserve all of the
sixteen supercharges present in the original theory, and are therefore
the most constrained.  On the other hand, reductions on curved
Ricci-flat manifolds, also produce manageable theories even though
there is little supersymmetry left.  The reason is that, as we will
review below, whatever supersymmetry remains becomes BRST-like,
rendering the theory cohomological.

The theory we will describe in what follows can be understood as that
arising out of euclidean D-branes wrapping around a Calabi--Yau
3-fold.  More prosaically, it is the dimensional reduction of
10-dimensional supersymmetric Yang--Mills theory to such a manifold.
Results in this direction for other manifolds have been obtained in
\cite{BSV,BTNT2too}, who considered euclidean D-branes wrapping around
calibrated submanifolds.  The resulting theories on the D-brane were
seen to be topologically twisted Yang--Mills theory -- the components
of the 10-dimensional gauge field in directions normal to the D-brane
being sections of the normal bundle to the calibrated submanifold
which need not be trivial.  In \cite{AFOS} the dimensional reductions
of supersymmetric Yang--Mills to 7- and 8-manifolds of exceptional
holonomy ($G_2$ and $\Spin(7)$, respectively) were studied.  The
theories obtained are cohomological \cite{WittenTFT} and localise on
the moduli space of generalised instantons and, in the 7-dimensional
case, monopoles.  The instanton theories agree (morally) with the
cohomological theories studied in \cite{BKS,AOS}.  Similar
considerations, in less detail but in more generality, can be found in
\cite{BTESYM}.

In this paper we will follow the approach of \cite{AFOS} and study
the theory on a Calabi--Yau 3-fold.  We will recover a cohomological
theory which localises on the moduli space of solutions of the
K\"ahler--Yang--Mills equations.  These equations have been studied by
Donaldson \cite{D} (for K\"ahler surfaces) and by Uhlenbeck--Yau
\cite{UY} (in complex dimension three and above), who show that they
are in one-to-one correspondence with stable holomorphic vector
bundles.  Cohomological theories which localise on this moduli space
have been discussed in \cite{BKS} as a reduction of the
eight-dimensional cohomological theories, and also briefly in
\cite{BTESYM}.

Our approach will be the following.  We start with 10-dimensional
supersymmetric Yang--Mills theory and reduce it to $6$-dimensional
euclidean space.  The resulting lagrangian can be promoted to any spin
6-manifold $M$ by simply covariantising the derivatives with respect
to the spin connection; but the supersymmetry transformations will
fail to be a symmetry of the action unless the spinorial parameters
are covariantly constant.  This requires that $M$ admit parallel
spinors, and that means that the holonomy group must be a subgroup of
$\SU(3)$.  If we want $M$ to be irreducible then the holonomy must be
$\SU(3)$.  Covariance of the supersymmetry algebra under the holonomy
group implies that the commutator of two supersymmetry transformations
with parallel spinors as parameters will result (on shell and up to
gauge transformations) in a translation by a parallel vector.  Since
for the irreducible manifolds we consider there are no such vectors,
the supersymmetry transformation is a BRST symmetry.  This general
argument shows that the resulting theory is cohomological.

This paper is organised as follows.  In Section 2 we discuss the
dimensional reduction of 10-dimensional supersymmetric Yang--Mills
theory to 6-di\-men\-sion\-al euclidean space.  In Section 3 we
specialise to the theory defined on a manifold of holonomy $\SU(3)$: a
Calabi--Yau 3-fold, and show that it is indeed cohomological.  In
Section 4 we rewrite the theory in the form of a balanced
cohomological field theory in the sense of \cite{DM} and \cite{BTNT2}.

For convenience we briefly summarise our spinor conventions here.
We use the Minkowski signature $(-1,1,1,\ldots,1)$.  The unitary
charge conjugation matrix for the Clifford algebra generators
$\gamma_\mu$ is specified, for given $\sigma_d, \sigma_t \in \{\pm
1\}$, by
\begin{equation}
C \gamma_\mu C^{-1} = \sigma_d\, \gamma_\mu^t\qquad\text{and}\qquad
C^t = \sigma_t\, C~.
\end{equation}
For the spinor representations of $\SO(3,1)$ we use notation along
the lines of Wess and Bagger \cite{WessBagger}, $\sigma^I =
(1,\sigma^i)$ with indices $\sigma^I_{a \Dot a}$ and $\Bar\sigma^I =
(-1,\sigma^i)$ obeying $\Bar\sigma^{I \Dot a a} = -
\epsilon^{ab}\epsilon^{\Dot a\Dot b} \sigma^I_{b\Dot b}$.  We choose
$\epsilon^{1 2} = 1$ and $\epsilon^{ab} \epsilon_{bc} = - \delta^a_c$.
The dot distinguishes between the two spinor representations, $\psi_a$
and $\Bar\psi^{\Dot a}$, for which the $\SO(3,1)$ generators are
\begin{equation}
\sigma^{IJ} \equiv \tfrac14 (\sigma^I\Bar\sigma^J -
\sigma^J\Bar\sigma^I)\qquad\text{and}\qquad
\Bar\sigma^{IJ} \equiv \tfrac14 (\Bar\sigma^I\sigma^J -
\Bar\sigma^J\sigma^I)~.
\end{equation}
It is straightforward to see that with $(\psi_{a})^\dagger \equiv
\Bar\psi_{\Dot a}$ and $(\Bar\psi^{\Dot a})^\dagger \equiv \psi^a$, we
have, e.g., $\Bar\psi^{\Dot a} = \epsilon^{\Dot a\Dot b}
\Bar\psi_{\Dot b}$.

\section{Dimensional reduction to six dimensions}

Our starting point is 10-di\-men\-sion\-al supersymmetric Yang--Mills
theory.  It can be formulated in terms of a Lie algebra valued gauge
field $A_M$ and a negative chirality Majorana--Weyl adjoint spinor
$\Psi$.  The Lie algebra is assumed to possess an invariant metric,
denoted $(-,-)$ or sometimes $\Tr$.  The lagrangian is then given by
\begin{equation}\label{eq:symlag}
\eL = - \tfrac14 (F_{MN},F^{MN}) + \tfrac{i}2 (\Bar\Psi, \Gamma^M D_M
\Psi)~,
\end{equation}
where $F_{MN} = \partial_M A_N - \partial_N A_M +  [A_M,A_N]$, and 
$D_M \Psi = \partial_M\Psi +  [A_M,\Psi]$, and
\begin{equation}
\Bar\Psi = \Psi^t \sC\, ,\quad\quad 
(\sigma_d^{(10)} = \sigma_t^{(10)} = -1)\, .
\end{equation}
This action is hermitian and invariant under the following
supersymmetry transformations,
\begin{equation}
\delta A_M = i \Bar\varepsilon \Gamma_M \Psi\qquad\text{and}\qquad
\delta \Psi = \tfrac12 F_{MN} \Gamma^{MN} \varepsilon~,
\end{equation}
where $\varepsilon$ is a constant negative chirality Majorana--Weyl
spinor.  The supersymmetry algebra only closes on-shell and up to
gauge transformations.

Reducing the theory down to six euclidean dimensions breaks the
10-di\-men\-sion\-al Lorentz invariance down to a subgroup
$\SO(3,1)\times\SO(6)$.  The first step of the dimensional reduction
is then the decomposition of our fields into irreducible
representations of this subgroup.  For tensor fields this is the
obvious decomposition $M = (I,\mu)$; in particular, $A_M = (\phi_I,
A_\mu)$.  For the spinors we use the representation
\begin{equation}
\Gamma^I = \Tilde\gamma^I \otimes \Bar\gamma_7 \qquad\text{and}\qquad
\Gamma^{\mu+4} = \1_4 \otimes \Bar\gamma^\mu~,
\end{equation}
where $\Bar\gamma^\mu$ are the generators for the Clifford algebra
$\Cl(6,0)$ and $\Tilde\gamma^I$ for $\Cl(3,1)$.  A straightforward
calculation gives
\begin{equation}
\Gamma_{11} = \Tilde\gamma_5 \otimes \Bar\gamma_7~.
\end{equation}
The charge conjugation matrix $\sC$ then decomposes as $\sC = \Tilde C
\otimes \Bar C$.  For definiteness we choose the representation in
which the $\Bar\gamma^\mu$ are all antisymmetric ($\sigma_d^{(6)} = -
\sigma_t^{(6)} = -1$), and the chiral representation for the
4-di\-men\-sion\-al $\gamma$'s ($\sigma_d^{(4)} = - \sigma_t^{(4)} =
1$); in terms of Pauli matrices,
\begin{align}
\Tilde\gamma^0 &= \1_2 \otimes (i \sigma^2) \\
\Tilde\gamma^i &= \sigma^i \otimes \sigma^1 \qquad\text{for
$i=1,2,3$}\\
\Tilde\gamma_5 &= \1_2 \otimes \sigma^3~,
\end{align}
with $\Tilde C = i\sigma^2 \otimes \sigma^3$.

Let $e_a$, $a=1,2$ denote an orthonormal eigen-basis of $\sigma^3$.
Then the Weyl condition determines
\begin{equation}
\Psi = e^a\otimes e_1\otimes\psi_{L a} + e_{\Dot a}\otimes e_2\otimes
\psi_R^{\Dot a}~.
\end{equation}
The 10-di\-men\-sion\-al Majorana condition then reduces to a reality
condition on the 6-di\-men\-sion\-al fields,
\begin{alignat}{2}
\Bar\psi_{L\Dot b} &= - \psi^{t \Dot a}_R \epsilon_{\Dot a\Dot b}&\qquad
\Bar\psi_R^b &= \psi_{L a}^t \epsilon^{ab} \\
\psi_{L a} &= \epsilon_{ab}\psi_R^{* b}&\qquad
\psi_R^{\Dot a} &= - \epsilon^{\Dot a\Dot b}\psi^*_{L\Dot b}~.
\end{alignat}
Finally then, the lagrangian reduces to
\begin{multline}
\eL =  -\tfrac14  \|F_{\mu\nu}\|^2 - \tfrac12 (D_\mu \phi_I, D_\mu
\phi^I) - \tfrac14  ([\phi_I,\phi_J],[\phi^I,\phi^J])\\
+ \tfrac{i}2 (\Bar\psi_R^a, \Bar\gamma_\mu D_\mu \psi_{La}) +
  \tfrac{i}2 (\Bar\psi_{L\Dot a}, \Bar\gamma_\mu D_\mu \psi_R^{\Dot
  a}) +  i (\Bar\psi_R^a, \sigma^I_{a\Dot b} [\phi_I,\psi_R^{\Dot
  b}])~,
\end{multline}
which is invariant under the supersymmetry transformations
\begin{align}
\delta \psi_{La} &= \tfrac12 \Bar\gamma_{\mu\nu} F_{\mu\nu}
\epsilon_{La} + \Bar\gamma_\mu D_\mu \psi_I
\sigma^I_{a\Dot b}\epsilon_R^{\Dot b} + [\phi_I,\phi_J]\sigma_a^{IJ b}
\epsilon_{Lb}\\
\delta \psi_R^{\Dot a} &= \tfrac12 \Bar\gamma_{\mu\nu} F_{\mu\nu}
\epsilon_R^{\Dot a} - \Bar\gamma_\mu D_\mu\psi_I \Bar\sigma^{I\Dot a
b} \epsilon_{L b} + [\phi_I,\phi_J] \Bar\sigma^{IJ\Dot a}{}_{\Dot b}
\epsilon_R^{\Dot b} \\
\delta A_\mu &= i \Bar\epsilon_R^a \Bar\gamma_\mu \psi_{L a} + i
\Bar\epsilon_{L\Dot a} \Bar\gamma\psi_R^{\Dot b}\\
\delta\phi_I &= -i \Bar\epsilon_{L\Dot a}\Bar\sigma_I^{\Dot a
b}\psi_{L b} + i \Bar\epsilon_R^a \sigma_{I a \Dot b}\psi_R^{\Dot
b}~.
\end{align}

\section{Reduction to manifolds with SU(3) holonomy}

Now that we have a supersymmetric theory defined on a six dimensional
euclidean space, it is time to extend it to a Calabi--Yau 3-fold.  The
structure group of the tangent bundle reduces to an $\SU(3)$ subgroup
of $\SO(6)$.  Our first task is to decompose the $\SO(6)$ fields into
irreducible representations of $\SU(3)$.  We will actually consider
the decomposition into $\U(3)$ irreducibles, $\U(3)$ being the
holonomy group of a 6-di\-men\-sion\-al K\"ahler manifold.  Since
$\U(3)$ is locally isomorphic to $\SU(3)\times\U(1)$, we will be able
to read off the $\SU(3)$ representations easily.  It is, of course,
sufficient to work in a local frame.

The embedding $\SO(6)\supset\SU(3)\times\U(1)$ leads to the branching
$\mathbf{4} = (\mathbf{1})_3 \oplus (\mathbf{3})_{-1}$; thus, under
the global symmetry $\SU(3)\times \U(1) \times \SO(3,1)$, we have that
the spinors $\lambda_R$ and $\lambda_L$ transform according to
\begin{align}
\lambda_R &\sim (\mathbf{1},3,\mathbf{2}_L) \oplus
(\mathbf{3},-1,\mathbf{2}_L) \\
\lambda_L &\sim (\mathbf{1}, -3,\mathbf{2}_R) \oplus
(\Bar{\mathbf{3}},1,\mathbf{2}_R)~.
\end{align}

Let $\theta$ denote the (commuting, left-handed) spinor which is
responsible for splitting the $\mathbf{4}$ above, and let us normalise
it to $\theta^\dagger \theta = 1$.  Clearly $\theta^*$ is the
right-handed singlet spinor, which splits the $\Bar{\mathbf{4}}$.  We
need the explicit projections onto these representations.  The
projector onto the singlet in the $\Bar{\mathbf{4}}$ is
\begin{equation}\label{eq:jmea}
\theta \theta^\dagger = \tfrac18 (1 - \Bar\gamma_7) - \tfrac1{16}
\theta^\dagger  \Bar\gamma_{\mu\nu}\theta (1 - \Bar\gamma_7)
\Bar\gamma_{\mu\nu}~,
\end{equation}
as follows from the standard result
\begin{equation}
\Bar\gamma^{\mu_1 \ldots \mu_r} = (-1)^{1 + r(r-1)/2} \frac{i}{(6-r)!}
\epsilon^{\mu_1\ldots \mu_6} \Bar\gamma_7 \Bar\gamma_{\mu_{r+1} \ldots
\mu_6}~,
\end{equation}
together with a Fierz transformation upon noticing that by chirality
\begin{equation}
\theta^\dagger \Bar\gamma^{(A)}\theta = 0~, \qquad\text{for $|A|$
odd.}
\end{equation}
It follows immediately from
\begin{equation}
\Bar\gamma^\lambda \Bar\gamma_{(A)} \Bar\gamma_\lambda = (-1)^{|A|} (6
- 2|A|)\, \Bar\gamma_{(A)}
\end{equation}
that
\begin{equation}\label{eq:jmeb}
\Bar\gamma_\lambda\theta \theta^\dagger\Bar\gamma_\lambda = \tfrac34
(1 + \Bar\gamma_7) - \tfrac18 \theta^\dagger \Bar\gamma_{\mu\nu}\theta
(1 + \Bar\gamma_7) \Bar\gamma_{\mu\nu}~,
\end{equation}
which will also be required later.

We may now introduce the K\"ahler form $k_{\mu\nu} \equiv i
\theta^\dagger \Bar\gamma_{\mu\nu}\theta$ and the 3-form
\begin{equation}\label{eq:jmed}
\Omega_{\mu\nu\lambda} \equiv \theta^\dagger\Bar\gamma_{\mu\nu\lambda}
\theta^*~.
\end{equation}
There are no other covariants since
\begin{equation}\label{eq:jmee}
\theta^\dagger\Bar\gamma^\mu \theta^* = 0~,
\end{equation}
which follows from sandwiching (the complex conjugate of)
\eqref{eq:jmeb} between $\theta^t$ and $\theta^*$.

At this point it is useful to make a special choice of $\theta$ which
corresponds to the standard choice of complex coordinates.  This
reduces the problem to the usual construction of the spinor
representation of $\SO(6)$ via linear combinations of the Clifford
algebra generators which obey the algebra of fermionic oscillators.
First introduce the combinations (taking $\mu = (\alpha, \Bar\alpha)$
in flat (local frame) coordinates)
\begin{equation}
\gamma^\alpha = \tfrac1{\sqrt{2}} (\Bar\gamma^\alpha + i
\Bar\gamma^{\alpha+3}) \qquad\text{and}\qquad
\gamma^{\Bar\alpha} = \tfrac1{\sqrt{2}} (\Bar\gamma^\alpha - i
\Bar\gamma^{\alpha+3})~.
\end{equation}
The $\SU(3)$ generators are then $T^\alpha{}_\beta =
\gamma^\alpha\gamma_\beta - \tfrac13 \delta^\alpha_\beta \gamma^\gamma
\gamma_\gamma$.  Requiring that $\theta$ be an $\SU(3)$ singlet (with
appropriate $U(1)$ charge $-3$) fixes $\gamma_\alpha \theta = 0$,
so that $k_{\alpha\beta} = k_{\Bar\alpha\Bar\beta} = 0$ and
$k_{\alpha\Bar\beta} = i \delta_{\alpha\Bar\beta}$.  Similarly all
components of $\Omega$ vanish by \eqref{eq:jmee} but for
$\Omega_{\alpha\beta\gamma} \equiv
\theta^\dagger\gamma_{\alpha\beta\gamma} \theta^*$ and its conjugate.
For completeness, notice that
\begin{equation}
\Omega_{\alpha\beta\gamma}
\Bar\Omega_{\Bar\alpha\Bar\beta'\Bar\gamma'} = 8
(\delta_{\beta\Bar\beta'} \delta_{\gamma\Bar\gamma'} -
\delta_{\beta\Bar\gamma'} \delta_{\gamma\Bar\beta '})~.
\end{equation}
Using vielbeins to translate to the coordinate basis, these results
apply for an arbitrary K\"ahler (6d) manifold.

The projectors for spinors onto $\SU(3)\times \U(1)$ covariant fields
follow directly by combining \eqref{eq:jmea} and (the complex
conjugate of) \eqref{eq:jmeb} to get the appropriate completeness
relations; e.g.,
\begin{equation}
\tfrac12(1-\Bar\gamma_7)=\theta\theta^\dagger + \tfrac12
\gamma_\alpha\theta^*\theta^t\gamma_{\Bar\alpha}~.
\end{equation}
For arbitrary symplectic Majorana--Weyl spinors $\chi_L$ or $\chi_R$,
define
\begin{equation}
\chi_a = \theta^\dagger \chi_{L a}\qquad\text{and}\qquad
\chi_{a\Bar\alpha} = \theta^t \gamma_{\Bar\alpha} \chi_{L a}~.
\end{equation}
Then the $\SO(3,1)$ covariant decompositions under $\SU(3)\times\U(1)$
are
\begin{equation}
\chi_{L a} = \theta \chi_a + \tfrac12 \gamma_\alpha\theta^*
\chi_{a\Bar\alpha}\qquad\text{and}\qquad
\chi_R^{\Dot a} = - \theta^* \epsilon^{\Dot a\Dot b}\Bar\chi_{\Dot b} + 
\tfrac12 \gamma_{\Bar\alpha} \theta \epsilon^{\Dot a\Dot b}
\Bar\chi_{\Dot b\alpha}~,
\end{equation}
and in terms of these fields, the lagrangian becomes
\begin{multline} \label{eq:jmmab}
\eL = -\tfrac12 (F_{\alpha\beta}, F_{\Bar\alpha\Bar\beta}) - \tfrac12
(F_{\alpha\Bar\beta}, F_{\Bar\alpha\beta}) - (D_\alpha\phi^I,
D_{\Bar\alpha} \phi_I)\\
- \tfrac14 ([\phi_I,\phi_J],[\phi^I,\phi^J]) + i \epsilon^{ab}
  (\psi_a, D_\alpha \psi_{b\Bar\alpha}) + i \epsilon^{\Dot a\Dot b}
  (\Bar\psi_{\Dot a}, D_{\Bar\alpha} \Bar\psi_{\Dot b\alpha})\\
- i  \Bar\sigma^{I\Dot a b} (\Bar\psi_{\Dot a}, [\phi_I,\psi_b]) -
  \tfrac{i}2 \Bar\sigma^{I\Dot a b} (\Bar\psi_{\Dot a \alpha},
  [\phi_I,\psi_{b\Bar\alpha}])\\
- \tfrac{i}8 \Omega_{\alpha\beta\gamma} \epsilon^{ab}
  (\psi_{a\Bar\alpha}, D_{\Bar\beta} \psi_{b\Bar\gamma}) - \tfrac{i}8
  \Bar\Omega_{\Bar\alpha\Bar\beta\Bar\gamma} \epsilon^{\Dot a\Dot b}
  (\Bar\psi_{\Dot a\alpha}, D_\beta \Bar\psi_{\Dot b\gamma})~.
\end{multline}
This action is invariant with respect to the supersymmetry
transformations with the parallel spinors as parameter.  These are
obtained from $\delta_S$ just by inserting the $\SU(3)$ singlet
grassmann parameters,
\begin{equation}
\epsilon_{L a} = \theta \epsilon_a~, \qquad 
\epsilon_R^{\Dot a} = - \theta^* \epsilon^{\Dot a\Dot b}
\Bar\epsilon_{\Dot b}~.
\end{equation}

In writing the explicit supersymmetries it is convenient to introduce
auxiliary fields so that the supersymmetry algebra closes off shell.
Because of $\SU(3)$ covariance and the fact that there are no
$\SU(3)$ invariant vectors on a Calabi--Yau 3-fold, the supersymmetry
algebra will be BRST-like, at least up to gauge transformations.
Further, it is convenient to split the conjugate generators using the
complex structure.   Thereto, we introduce supercharges via
\begin{equation}
\delta_S = i\Bar\epsilon_{\Dot a} \Bar Q^{\Dot a} - i \epsilon_a Q^a~.
\end{equation}
The sign is such that $Q$ and $\Bar Q$ act like canonical generators
(so $Q A = B \Rightarrow \Bar Q A^\dagger = -B^\dagger$ and
$Q B = A \Rightarrow \Bar Q B^\dagger = A^\dagger$, if $A$ is bosonic).

The algebra of charges on the fields is then easily found to be 
\begin{equation}
\{Q^a,Q^b\} = 0~,\quad\{Q^{\Dot a},Q^{\Dot b}\} =
0\quad\text{and}\quad \{Q^a,\Bar Q^{\Dot b}\} = \delta_G(-2i
\Bar\sigma^{I\Dot b a}\phi_I)~,
\end{equation}
where $\delta_G(\theta)$ means ``gauge transformation with parameter
$\theta$''. Then $\delta_S$ can be extended to the auxiliary fields 
\begin{equation}
H = - i F_{\alpha\Bar\alpha}\qquad\text{and}\qquad
H_\alpha = \tfrac{i}2 \Omega_{\alpha\beta\gamma}
F_{\Bar\beta\Bar\gamma}~,
\end{equation}
so that the supersymmetry algebra is maintained off shell.  We have
the explicit transformations:

\begin{center}
\renewcommand{\arraystretch}{1.2}
\begin{tabular}{|>{$}c<{$}|>{$}c<{$}|>{$}c<{$}|}\hline
\text{Field} & \Bar Q^{\Dot a} & Q^a \\
\hline
\hline
\psi_b & 0 & H\delta^a_b + i \sigma^{IJa}{}_b [\phi_I,\phi_J]\\
\Bar\psi_{\Dot b} & H\delta^{\Dot a}_{\Dot b} + i
\Bar\sigma^{IJ\Dot a}{}_{\Dot b} [\phi_I,\phi_J] & 0 \\
\psi_{b\Bar\alpha} & 2i D_{\Bar\alpha} \phi_I \Bar\sigma^{I\Dot a a}
\epsilon_{ab} & H_{\Bar\alpha} \delta^a_b \\ 
\Bar\psi_{\Dot b \alpha} & H_{\alpha} \delta^{\Dot a}_{\Dot b} & 2i
D_\alpha\phi_I \epsilon^{ab}\sigma^I_{b\Dot b} \\
A_\alpha & \epsilon^{\Dot a\Dot b}\Bar\psi_{\Dot b \alpha} & 0 \\
A_{\Bar\alpha} & 0 &  -\epsilon^{ab}\psi_{b\Bar\alpha} \\
\phi_I & - \Bar\sigma_I^{\Dot a b} \psi_b & \Bar\sigma_I^{\Dot b a}
\Bar\psi_{\Dot b} \\
H & i \Bar\sigma^{I\Dot a b}[\phi_I,\psi_b] & i \Bar\sigma^{I\Dot b a}
[\phi_I,\Bar\psi_{\Dot b}] \\ 
H_\alpha & 0 & 4i \epsilon^{ab}D_\alpha\psi_b + 2i\Bar\sigma^{I\Dot b
a} [\phi_I,\Bar\psi_{\Dot b\alpha}] \\
H_{\Bar\alpha} & 4i \epsilon^{\Dot a\Dot b}
D_{\Bar\alpha}\Bar\psi_{\Dot b} + 2i\Bar\sigma^{I\Dot a b}
[\phi_I,\psi_{b\Bar\alpha}] & 0\\
\hline
\end{tabular}
\end{center}

It is possible now to reduce to a cohomological theory with a single
cohomological symmetry: setting $\psi_2 = \Bar \psi_{\Dot 2} = 0$,
which requires $\phi_1 = \phi_2 = 0$, we are left with the
supersymmetries generated by $Q^1$ and $\Bar Q^{\Dot 1}$.  Instead we
will keep all supersymmetries and work out a balanced formulation for
this cohomological theory.

\section{A balanced cohomological field theory}

In order to recognise what this theory computes, it will prove
convenient to rewrite it in balanced form \cite{BTNT2,DM}; that is, in
terms of potentials.  Let us first write the lagrangian in a form
linear in $Q$'s.  To this effect, introduce
\begin{equation}
\Bar\eL = Q^a \eV_a + \Bar Q^{\Dot a} \Bar \eV_{\Dot a} \, ,
\end{equation}
where $\eV_a$ is dimension $\tfrac72$ in the natural units where the
gauge coupling is scaled out, $A_\mu$ and $\phi_I$ have dimension 1
and $\psi$'s have dimension $\tfrac32$.  Further, it should be gauge
invariant and an $\SO(3,1)$ doublet.  Taking the most general possible
Ansatz and comparing to \eqref{eq:jmmab}, we find
\begin{multline}
\eV_a = \tfrac{i}4  (\psi_a, F_{\alpha\Bar\alpha}) + \tfrac18 (\psi_a,
H) - \tfrac{i}8 \sigma^{IJb}{}_a (\psi_b, [\phi_I,\phi_J]) -
     \tfrac{i}{16} \Omega_{\alpha\beta\gamma} (\psi_{a\Bar\alpha},
     F_{\Bar\beta\Bar\gamma})\\
+ \tfrac1{16} (\psi_{a\Bar\alpha}, H_\alpha) - \tfrac{i}8
  \epsilon^{\Dot a\Dot b} \sigma^I_{a\Dot a} (\Bar\psi_{\Dot b
  \alpha}, D_{\Bar\alpha}\phi_I)~.
\end{multline}
Eliminating the auxiliary fields (which are determined correctly),
we find that
\begin{equation}\label{eq:topterm}
\eL = \Bar \eL + \tfrac12 (F_{\alpha\beta}, F_{\Bar\alpha\Bar\beta}) -
\tfrac12 (F_{\alpha\Bar\beta}, F_{\Bar\alpha\beta}) - \tfrac12
(F_{\alpha\Bar\alpha}, F_{\beta\Bar\beta})~.
\end{equation}
Note that the extra terms can be rewritten as $-\tfrac12  k \wedge \Tr
(F \wedge F)$, whence their integral only depends on the K\"ahler
class and the characteristic class of the gauge bundle.

We can pursue this a little further, writing $\Bar \eL$ quadratic in
$Q$'s.  The most general form is
\begin{equation}
\Bar \eL = \epsilon_{ab} Q^a Q^b \eV + Q^a \sigma^I_{a \Dot a} \Bar
Q^{\Dot a} \eV_I + \text{h.c.}~,
\end{equation}
and a similar analysis to the above gives
\begin{equation}
\eV_a = \epsilon_{ab} Q^b \eV + \sigma^I_{a \Dot a} \Bar Q^{\Dot a}
\eV_I~,
\end{equation}
where
\begin{align}
\eV &= - \tfrac{i}{32} \Omega_{\alpha\beta\gamma}
\CS(A)_{\Bar\alpha\Bar\beta\Bar\gamma} -\tfrac1{16} \epsilon^{cd}
(\psi_c,\psi_d) \\
\eV_I &= - \tfrac{i}{16} (\phi_I, F_{\alpha\Bar\alpha}) + \tfrac1{64}
\Bar\sigma_I^{\Dot b b} (\Bar\psi_{\Dot b \alpha}, \psi_{b
\Bar\alpha})~,
\end{align}
with $\CS(A)$ the holomorphic Chern--Simons 3-form,
\begin{equation}
\CS(A)_{\Bar\alpha\Bar\beta\Bar\gamma} = (A_{\Bar\alpha},
F_{\Bar\beta\Bar\gamma}) - \tfrac13 (A_{\Bar\alpha}, [A_{\Bar\beta},
A_{\Bar\gamma}])~.
\end{equation}
Clearly
\begin{equation}\label{eq:twostars}
\Bar Q^{\Dot a} \eV = 0~,
\end{equation}
and $\eV_I$ is real.  Thus we can write
\begin{equation}
\Bar \eL \equiv \Bar\eL_1 + \Bar\eL_2 = (\epsilon_{ab} Q^a Q^b -
\epsilon_{\Dot a\Dot b} \Bar Q^{\Dot a} \Bar Q^{\Dot b}) (\eV + \Bar
\eV) +  2 Q^a \sigma^I_{a \Dot a} \Bar Q^{\Dot a} \eV_I
\end{equation}
Note that the holomorphic Chern--Simons term cannot be reproduced as a
BRST variation, so it isn't profitable to continue this process.  That
such a term---only invariant under small gauge
transformations---should appear at all is quite interesting, and
consistent with the results in \cite{AFOS} for manifolds of $G_2$
holonomy.

We would like to rewrite this in balanced form along the lines of
\cite{DM} (see also \cite{BTNT2}).  To do that, one must first choose
a global $\SL(2,\RR)$ under which the balanced supercharges will
transform as a doublet $d$.  The lagrangian must then be written (up
to a topological term) in the form
\begin{equation}\label{eq:balancedform}
\epsilon_{AB} d^A d^B \eW\quad\text{for some potential $\eW$,}
\end{equation}
where the critical points of $\eW$ agree with the fixed points
of the cohomological symmetry.

It is natural, in our case, to take the $\SO(2,1)$ subgroup of the
global $\SO(3,1)$ symmetry.  Then the doublet supercharges can be
taken as the linear combinations $d^A$ and $\Tilde d^A$, where
\begin{equation}
d = \begin{pmatrix}
\Bar Q^{\Dot 1} + Q^2 \\ \Bar Q^{\Dot 2} + Q^1
\end{pmatrix}~,
\qquad 
\Tilde d = \begin{pmatrix} \Bar Q^{\Dot 1} - Q^2 \\ \Bar Q^{\Dot 2} -
Q^1
\end{pmatrix}~,
\end{equation}
and $\Bar S$ can be decomposed.  The first term reduces to
\begin{equation}
\Bar \eL_1 = -\tfrac12 (\epsilon_{AB} d^A d^B + \epsilon_{AB} \Tilde
d^A \Tilde d^B)(\eV+\Bar \eV)~,
\end{equation}
while the second term is just
\begin{equation}
\Bar \eL_2 = 2 Q^a \sigma^3_{a \Dot a} \Bar Q^{\Dot a} \eV_3 + 2 Q^a
\sigma^\mu_{a \Dot a} \Bar Q^{\Dot a} \eV_\mu~.
\end{equation}
Remarkably, an explicit calculation shows that both of the terms in
$\Bar \eL_2$ are individually $\SO(3,1)$ invariant.  Since there is a
unique such invariant bilinear in $Q^a$ and $\Bar Q^{\Dot a}$ these
two terms in $\Bar \eL_2$ must be proportional, and we can consider
just the first.  Thus $\Bar \eL_2$ is itself proportional to
\begin{equation}
2 Q^a \sigma^3_{a \Dot a} \Bar Q^{\Dot a} \eV_3 = -\tfrac12
(\epsilon_{AB} d^A d^B - \epsilon_{AB} \Tilde d^A \Tilde d^B) \eV_3~.
\end{equation}

This is still not quite in balanced form, since we have two doublets
of supercharges.  However, it is straightforward to check that
\begin{equation}
\epsilon_{AB} d^A d^B(\eV+\Bar \eV) = \epsilon_{AB} \Tilde d^A \Tilde
d^B (\eV+\Bar \eV)~,
\end{equation}
since by \eqref{eq:twostars} the difference may be written as
anticommutators of supercharges.  Moreover, another explicit
calculation shows that
\begin{equation}
\epsilon_{AB} d^A d^B \eV_3 = - \epsilon_{AB} \Tilde d^A \Tilde d^B
\eV_3~.
\end{equation}

Collecting these results we see that $\Bar\eL$ is of balanced form
\eqref{eq:balancedform} with potential,
\begin{multline}
\eW = \tfrac{i}4 (\varphi, F_{\alpha\Bar\alpha}) + \tfrac{i}{32}
\Bar\Omega_{\Bar\alpha\Bar\beta\Bar\gamma} \CS(A)_{\alpha\beta\gamma}
- \tfrac{i}{32} \Omega_{\alpha\beta\gamma}
  \CS(A)_{\Bar\alpha\Bar\beta\Bar\gamma} \\
- \tfrac1{16} (\Bar\psi_{\Dot a \alpha}, \psi_{a\Bar\alpha}) +
  \tfrac1{16} \epsilon^{ab} (\psi_{a}, \psi_{b}) - \tfrac1{16}
  \epsilon^{\Dot a\Dot b} (\Bar \psi_{\Dot a}, \Bar \psi_{\Dot b})~,
\end{multline}
where we have introduced $\varphi\equiv\phi_3$.  Hence the physical
supersymmetric Yang--Mills theory differs from this balanced 
cohomological field theory by the topological term $T$ in
\eqref{eq:topterm}.  Note, however, that $T$ of course depends on the
K\"ahler class.

Balanced theories localise on the critical points of $\eW$.  These
points correspond to fermions set to zero, and bosons obeying
$F_{\alpha\Bar\alpha} = 0$ and the Bogomol'nyi-type equations
\begin{equation}
\tfrac14 \Omega_{\alpha\beta\gamma} F_{\Bar\beta\Bar\gamma} +
D_{\Bar\alpha} \varphi = 0 \qquad\text{and}\qquad
\tfrac14 \Omega_{\Bar\alpha\Bar\beta\Bar\gamma} F_{\beta\gamma} +
D_{\alpha} \varphi = 0~.
\end{equation}
For a compact Calabi--Yau 3-fold, these equations reduce to the
K\"ahler--Yang--Mills equations
\begin{equation}
F_{\alpha\beta} = F_{\Bar\alpha\Bar\beta} = 0 \qquad\text{and}\qquad
F_{\alpha\Bar\alpha} = 0~,
\end{equation}
together with the trivial $D_\alpha\varphi = D_{\Bar\alpha}\varphi =
0$.

\section{Conclusions and Outlook}

We have shown that the dimensional reduction of supersymmetric
Yang--Mills on a compact Calabi--Yau 3-fold is a cohomological theory
which localises on the moduli space of solutions to the K\"ahler
Yang--Mills equations or, by the work of Donaldson and Uhlenbeck--Yau,
on the moduli space of stable holomorphic bundles.  Observables in
this theory correspond to invariants of this moduli space, which
generalise the Donaldson invariants in four dimensions.  Unlike four
dimensions, these are not topological invariants of the Calabi--Yau
3-fold, but a priori only invariants of the $\SU(3)$ structure.  It
follows from the balanced formulation \eqref{eq:balancedform} of the
theory that $\Bar\eL$ is invariant under infinitesimal deformations of
the metric which preserve the Calabi--Yau condition.  A similar result
was shown in \cite{AOS} for the cohomological theories on 7- and
8-manifolds of exceptional holonomy.

One direction in which this work may be pursued is to examine the
Ansatz of \cite{BSV} for the effective theory of branes wrapped around
supersymmetric cycles in the Calabi-Yau space, here a special
lagrangian torus.  Considering the embedding of the torus local
coordinates of \cite{SYZ} to see how the topological twisting on the
torus arises, we note that arguments as in \cite{BSV} (and
\cite{BJSV}) suggest that, for one $\U(1)$ case, the resulting path
integral on the torus localises on the moduli space $\eM_{\mathrm{SL}}
\times \eM_{\mathrm{Flat}}$, precisely the local description of the
mirror \cite{SYZ}.  Details will appear elsewhere.

\section*{Acknowledgement}

JMF takes pleasure in thanking Bobby Acharya, Chris Hull, Chris K\"ohl
and Bill Spence for conversations on this and other related topics.

\end{document}